\documentclass[journal]{IEEEtran}
\usepackage[T1]{fontenc}

\usepackage[noadjust]{cite} 

\ifCLASSINFOpdf
\usepackage[pdftex]{graphicx}
\usepackage{epstopdf}
\usepackage{mwe}
\usepackage{caption}
\usepackage{subcaption}
\usepackage[justification=centering]{caption}
\usepackage[justification=centering]{subcaption}
\else
\fi
\usepackage{amsmath}
\usepackage{mathrsfs,amssymb}

\usepackage{commath}
\usepackage{amssymb}
\DeclareMathAlphabet{\mathpzc}{OT1}{pzc}{m}{it}
\usepackage{amsthm}
\usepackage{thmtools,thm-restate}
\newtheorem{lemma}{Lemma}
\usepackage[cmintegrals]{newtxmath}
\usepackage{algorithm}
\usepackage{algpseudocode}
\algnewcommand{\IfThenElse}[3]{
  \State \algorithmicif\ #1\ \algorithmicthen\ #2\ \algorithmicelse\ #3}
\makeatletter
\let\OldStatex\Statex
\renewcommand{\Statex}[1][3]{%
  \setlength\@tempdima{\algorithmicindent}%
  \OldStatex\hskip\dimexpr#1\@tempdima\relax}
\makeatother
\begin{document}
%
\title{On Green Multicasting over Cognitive Radio Fading Channels}
%
%
%

\author{Sangeeta~Bhattacharjee,~\IEEEmembership{Student Member,~IEEE,}
      Tamaghna~Acharya,~\IEEEmembership{Member,~IEEE,}
       and~Uma~Bhattacharya 
\thanks{Sangeeta Bhattacharjee, Tamaghna Acharya are with the Department of Electronics and Telecommunication Engineering and Uma Bhattacharya is with the Department of Computer Science and Technology, Indian Institute of Engineering Science and Technology, Shibpur, Howrah, India, 711103 (e-mail: sangeeta.bhatta@gmail.com; t\textunderscore acharya@telecom.iiests.ac.in; ub@cs.iiests.ac.in).}}
\maketitle
\begin{abstract}
In this paper, an underlay cognitive radio (CR) multicast network, consisting of a cognitive base station (CBS) and multiple multicast groups of secondary users (SUs), is considered. All SUs, belonging to a particular multicast group, are served by the CBS using a common primary user (PU) channel. The goal is to maximize the energy efficiency (EE) of the system, through dynamic adaptation of target rate and transmit power for each multicast group, under the PUs' individual interference constraints. The optimization problem formulated for this is proved to be non quasi-concave with respect to the joint variation of the CBS's transmit power and target rate. An efficient iterative algorithm for EE maximization is proposed along with its complexity analysis.  Simulation results illustrate the performance gain of our proposed scheme.
\end{abstract}
\IEEEpeerreviewmaketitle
\section{Introduction}
\IEEEPARstart{E}{nergy-efficient} cognitive radio network (CRN) \cite{gur2011green} is envisaged as a promising paradigm for the design of next generation wireless networks, owing to its unique potential to achieve groundbreaking improvement in spectral efficiency in an energy efficient manner. An excellent review of energy efficiency (EE) maximizing approaches, covering all three possible modes of operation (i.e. underlay, interweave, overlay) in CRN may be found in \cite{huang2015green}. In this paper, we focus on the underlay mode in CRNs owing to its simplicity. Recently, a few more power control schemes are proposed to increase EE in underlay CRNs \cite{sboui2015energy,zhou2016energy} with fading channels. In \cite{sboui2015energy}, the authors propose energy efficient power allocation in CRN under peak and average power constraints of secondary users (SUs) and analyze the importance of the availability of channel state information (CSI) on EE of the network. 
In \cite{zhou2016energy}, power allocation schemes are proposed to maximize EE in CRNs carrying both delay-insensitive and delay-sensitive traffic. Multiple antenna aided energy efficient transmission in underlay CRN is investigated in \cite{zappone2015energy} and references therein.

Multicasting in CRNs \cite{qadir2014multicasting} is expected to further improve the spectral efficiency of such networks by enabling usage of a single primary user (PU) channel for supporting simultaneous downlink transmissions of a common message, by a cognitive base station (CBS), to a multitude of SUs. Further, it may also be feasible to support multimedia services, with relaxed quality of service (QoS) constraints, to SU multicast groups if the shared PU channels are wideband in nature. However, multicasting over CRN is even more challenging because a common minimum throughput is required to be delivered to each SU in a multicast group over the same PU channel without violating the stringent interference temperature (IT) constraint of the PU. Hence, an efficient transmit power adaptation scheme for multicast in a CRN should consider the channel conditions for CBS-PU interference link as well as CBS-SU link for each SU in the network. Xu et al. study minimization of group outage probability as well as weighted sum of individual outage probability  for multiple simultaneous multicast sessions in an underlay CRN, through optimal transmit power control for each group \cite{xu2011outage}. Here, the rate of transmission to any multicast group is assumed to be determined by the worst channel (CBS-SU) conditions among all the users in the group. The same authors, in their attempt to maximize average sum rate of all the multicast transmissions, suggest a more efficient approach of adapting the transmission rate along with the control of transmit power under the constraints of a probabilistic QoS for individual SUs in the group, average transmit power and average interference \cite{xu2013joint}.

In this paper, we consider a green CR multicast network, consisting of a CBS and multiple SU multicast groups. The CBS transmits data to several SU multicast groups simultaneously. The PU channels, necessary to support these transmissions, are assumed to be pre-assigned following the principle of one channel per multicast group and without any provision of reuse any PU channel. We consider the presence of a rate adaptive application layer along with a power adaptive link layer at the CBS. The proposed cross layer approach attempts to explore the possible interaction between these two layers to maximize EE of the CBS while supporting multiple multicast groups, with heterogeneous service requirements. We present a non linear optimization framework for maximizing the EE of the multicast CRN under PU's interference power threshold (IPT) constraint. Due to the non quasi-concave nature of the problem, an algorithm is designed based on the principle of coordinate descent method \cite{bertsekas1999nonlinear}. The proposed scheme determines the optimal value of transmit power and target rate for each multicast group, while maximizing the system EE. 

The rest of this paper is organized as follows. The system model is presented in section II. In section III, we formulate the EE maximization problem under IPT constraint and present an iterative algorithm.  Simulation results and related discussions are presented in section IV. Finally, section V concludes the paper.
\section{System Model and Preliminaries}
\begin{figure} [h]
\includegraphics[scale=0.45]{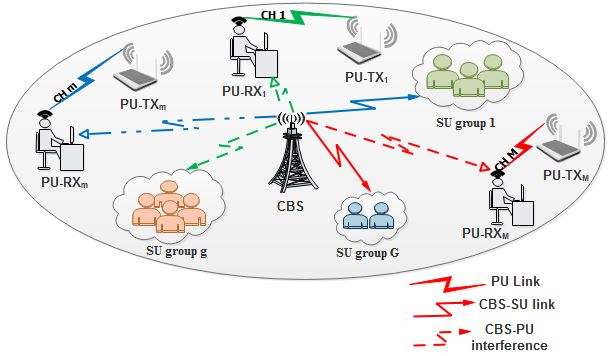} 
\centering
\caption{System Model}  \label{f1}
\end{figure}
The spectrum of interest is assumed to be divided into $M$ orthogonal channels which are licensed to $M$ independent PUs. As shown in Fig. \ref{f1}, each primary link comprises of one PU transmitter (PU-TX) and one PU receiver (PU-RX). A CRN with underlay mode of operation is considered, where a CBS serves $G$ multicast groups simultaneously, $G \leq M$. A multicast group, $g$, is assumed to be formed by a set of co-located SUs, $K_g$, enjoying a common service from the CBS ($g \in \left\lbrace1,2,....,G\right\rbrace $). Each SU node belongs to only one group during the multicast session. It is assumed that CBS can access all the $M$ PU channels and each SU group is exclusively assigned only one channel. Since our focus is to design a solution based on joint rate and power allocation, we assume that channel allocation and admission control mechanisms for $G$ multicast groups using $M$ PU channels are done by the CBS in advance. All the wireless channels that exist between the CBS and SUs of a multicast group are considered to be independent and identically distributed (i.i.d) block fading channels.\par
The instantaneous channel power gains for CBS to SU links at a particular fading state are denoted by $h_{g,k}^{ss}$, $k\in K_g$. The channel power gains on the interference links from CBS to PU-RXs are denoted by $h_{m,g}^{sp}$, where group $g$ is assigned $m^{th}$ PU channel. For ease of analysis, interferences from PU-TXs to SUs are represented by noise like term following \cite{xu2011outage}. Total interference plus noise at SUs is modelled as i.i.d, circularly symmetric complex Gaussian random variable with zero mean and variance $N_0$, denoted by $\mathcal{CN}(0,N_0)$. The instantaneous CSI from CBS to each SU of a multicast group is estimated at the SUs. Similar to \cite{zhou2016energy}, we assume that CBS is aware of the channel fading statistics for each multicast group, while the CSI of the interference links is obtained at CBS from a third party such as manager centre. This information is exploited by the CBS, for adapting its application layer rate and transmit power, to support energy efficient operation of the network. We also assume that the CBS disseminates the adapted rate values to all the SUs of a multicast group before transmission. Let $P_{g}$ denote the transmit power allocated by CBS to group $g$.
\subsection{Outage probability of multicast group} \label{s1}
The instantaneous rate that can be achieved by the SU $k{\in}K_g$ is given by
\begin{equation} \label{m2}
r_{g,k}=\log_2\left(1+\frac{h_{g,k}^{ss}P_{g}}{N_0}\right), \ g\in 1,2, \ldots, G
\vspace{-3mm}
\end{equation}
Since all the SUs in a multicast group are served using the same PU channel, the multicast capacity depends on the instantaneous link condition of each SU in the group. Hence, to ensure that all the SUs of a multicast group decode the received bits correctly, the transmission rate of CBS for a given multicast group $g$ is defined as the maximum achievable rate of the SU experiencing the worst channel condition in the group $g$ \cite{xu2011outage}. Thus, for a given CBS target rate, the outage probability for a multicast group is derived in Lemma 1.
\begin{lemma} \label{L1}
For any multicast group $g$, ($g\in 1,2, \cdots, G$), if CBS targets to support an application with rate $R_g$, assuming i.i.d channel condition between the CBS and each SU member of the multicast group, the outage probability of multicast group $g$ is given by
\begin{align}\label{m3}
\begin{split}
\mathscr{P}_{g}^{out}\left( P_g,R_g\right) &= Pr\left(\min\left\lbrace r_{g,k} \right\rbrace\leq R_g \right), \forall{k\in K_g}
\\
&=1-\left[ Pr\left(r_{g,k} > R_g \right)\right] ^{|K_g|}
\\
&=1-e^{-\frac{No|K_g|}{\lambda_{g}^{ss}P_g}(2^{R_g}-1)}
\end{split}
\end{align}
\normalfont{where $\lambda_{g}^{ss}$ is the mean of exponentially distributed channel power gain from CBS to SUs and $Pr(.)$ denotes the probability.}\\

\textit{Proof:} See Appendix 1.
\end{lemma}
\subsection{Energy Efficiency} \label{s2}
In fading channels, EE is defined as the ratio of average throughput to the average power consumed. Thus, the system EE is given by the average number of transmitted bits which may be reliably communicated over the channel per unit of energy consumed. Average throughput for multicast group $g$ may be written as
\vspace*{-3mm}
\begin{equation} \label{m4}
T_g=R_g\left( 1-\mathscr{P}_{g}^{out}\left( P_g,R_g\right)\right) 
\end{equation}
Using (\ref{m3}), we can rewrite (\ref{m4}) as
\begin{equation} \label{m5}
T_g=R_g*e^{-\frac{No|K_g|}{\lambda_{g}^{ss}P_g}(2^{R_g}-1)}
\end{equation}
Hence, EE of the CR multicast system is given by
\vspace*{-1mm}
\begin{align} \label{m6}
\begin{split}
\eta(P_g,R_g) &=\frac{\sum_{g=1}^{G}{R_g\left( 1-\mathscr{P}_{g}^{out}\right)}}{\sum_{g=1}^{G}{P_g+P_c}}
\\
&=\frac{\sum_{g=1}^{G}R_g*e^{-\frac{No|K_g|}{\lambda_{g}^{ss}P_g}(2^{R_g}-1)}}{\sum_{g=1}^{G}{P_g+P_c}}
\end{split}
\end{align}
\subsection{System Constraints}
\subsubsection{Interference Power Threshold (IPT)} \label{s3}
To ensure that PUs' communication remains stable, interferences caused by CBS transmission to group $g$ in $m^{th}$ PU channel, is limited by IPT constraint \cite{zhou2016energy} as follows:
\begin{equation} \label{m7}
\lambda_{m,g}^{sp}P_g\leq Q_m, \ \forall{m\in M}
\end{equation}
Here, $Q_m$ is the prescribed average interference threshold for $m^{th}$ PU. Further, $\lambda_{m,g}^{sp}$ 
denotes the mean of the exponentially distributed channel power gain from CBS to $m^{th}$ PU-RX.

\subsubsection{Maximum Outage Probability Constraint} \label{s3a}
To ensure QoS of multicasting, CBS must ensure that a minimum application layer rate $R_g^{min}$ is supported for each multicast group. The worst case outage performance of the application, to support multicast services to each group, with rate $R_g^{min}$ is limited by a predefined value, $\mathscr{P}_g^{out^{(max)}}$.\par. 
\section{Energy Efficient Joint Rate and Power allocation}
In this section, we propose a joint rate and power control scheme for CBS to maximize its EE while satisfying PUs' individual IPT constraints. The EE maximization problem for the multicast CRN is formulated in \textbf{P1}. 
\begin{equation} \label{m8}
\begin{split}
\textbf{P1}: \ \  \ \ \ \ \ \ \ \ \ \ \ \ \ \ \ \ \ &\max \eta(P_g,R_g)\\
s.t. \ \ (\ref{m7}), \ \ R_g^{min}\leq R_g&\leq R_g^{max},\ \ P_g^{min}\leq P_g\leq P_g^{max} \\
\text{where} \ \mathscr{P}_{g}^{out}\left( P_g^{min},R_g^{min}\right)&=\mathscr{P}_g^{out^{(max)}}, \ \forall{g\in 1,2, \cdots ,G},
\end{split}
\end{equation}
where $R_g^{min}$ and $R_g^{max}$ are respectively the minimum and maximum rate supported by the target application. $P_g^{max}$ is the maximum power that satisfies (\ref{m7}). $P_g^{min}$ is the transmit power required to support the application with rate $R_g^{min}$ and maximum outage, $\mathscr{P}_g^{out^{(max)}}$.\par

The EE function $\eta$, in \textbf{P1}, is not a quasi-concave function w.r.t $P_g$ while it is quasi-concave w.r.t $R_g$, as proved in the following subsections. Hence, applying standard fractional programming techniques to this problem is difficult. To solve this problem, we first investigate the variation of $\eta$ with each optimization variable and find its optimal value, while assuming the other variable as constant. We then propose an algorithm to maximize EE of the multicast CRN while satisfying IPT constraint.
\vspace*{-2mm}
\subsection{Maximize $\eta$ w.r.t $P_g$} \label{s4}
Considering $\eta(P_g,R_g)$ as a function of $P_g$ only, and treating all other variables as constants, the $1^{st}$ order derivative of (\ref{m6}) w.r.t $P_g$ is
\begin{equation}\label{m10}
\eta^{(1)}(P_g)=\frac{\mathbb{F}_1(P_g)}{(P_g+P_c)^2}
\vspace{-3mm}
\end{equation}
\begin{align}\label{m11}
\hspace*{-1.3cm} \text{where} \ \ &\mathbb{F}_1(P_g)=\frac{\left\lbrace -\chi P_g^2+\gamma P_g+\mu\right\rbrace}{P_g^2}e^{-\frac{\omega}{P_g}} -\zeta ,
\end{align}
\begin{align*}
\begin{split}
&\omega=\frac{No|K_g|}{\lambda_{g}^{ss}}(2^{R_g}-1), \ \chi=R_g, \ \gamma=\omega\chi, \\
&\mu=\gamma(\sum_{\substack{i=1\\ i\neq g}}^{G}P_i + P_c), \ \zeta=\sum_{\substack{i=1\\ i\neq g}}^{G}T_i
\end{split}
\end{align*}
From (\ref{m10}), it can be observed that sign of $\eta^{(1)}(P_g)$ is same as that of $\mathbb{F}_1(P_g)$. Also, from (\ref{m11}), it is clear that the sign of $\mathbb{F}_1(P_g)$ depends on following
\begin{equation}\label{m12}
\mathbb{F}_2(P_g)=-\chi P_g^2+\gamma P_g+\mu-\zeta P_g^2e^{\frac{\omega}{P_g}}
\end{equation}
Now, to determine the sign of $\mathbb{F}_2(P_g)$, we evaluate its $1^{st}$ and $2^{nd}$ order derivatives w.r.t $P_g$ as follows
\begin{equation}\label{m13}
\mathbb{F}_2^{(1)}(P_g)=-2\chi P_g+\gamma +\zeta e^{\frac{\omega}{P_g}}(\omega-2P_g)
\end{equation}
\begin{equation}\label{m14}
\mathbb{F}_2^{(2)}(P_g)=-2\chi-\zeta \scalebox{1}{$\left[\frac{P_g^2+(P_g-\omega)^2}{P_g^2}e^{\frac{\omega}{P_g}}\right]$} 
\end{equation}
From (\ref{m14}), it is seen that $\mathbb{F}_2^{(2)}$ is negative which implies $\mathbb{F}_2^{(1)}$ is a decreasing function of $P_g$.
Since, $\mathbb{F}_2^{(1)}(P_g)$ is continuous in $(0,\infty)$, while $\lim\limits_{P_g\to 0} \mathbb{F}_2^{(1)}(P_g)=\infty$ and $\lim\limits_{P_g\to \infty} \mathbb{F}_2^{(1)}(P_g)=-\infty$, it can be concluded that there exists a unique point $\beta$ such that $\mathbb{F}_2^{(1)}(\beta)=0$. In other words, $\mathbb{F}_2(P_g)$ is strictly increasing in $(0,\beta)$ and strictly decreasing in $(\beta,\infty)$. Also, $\lim\limits_{P_g\to 0} \mathbb{F}_2(P_g)=-\infty$ and $\lim\limits_{P_g\to \infty} \mathbb{F}_2(P_g)=-\infty$. Thus, depending on the sign of $\mathbb{F}_2(P_g)$, we may consider following two cases:
\begin{itemize}
\setlength{\itemindent}{0.1in}
\item [\textbf{(i)}]
If $\max{\mathbb{F}_2(P_g)}=\mathbb{F}_2(\beta)\leq 0 \ \forall P_g\in (0,\infty)$, it implies that $\eta(P_g)$ is a decreasing function of $P_g$.
\item [\textbf{(ii)}]
If $\max{\mathbb{F}_2(P_g)}=\mathbb{F}_2(\beta)> 0$, there exists two points $P_g=\alpha_1$ and $P_g=\alpha_2$ satisfying $\mathbb{F}_2(P_g)=0$, $0<\alpha_1<\beta<\alpha_2$. 
\end{itemize}
Thus, following sub-cases under \textbf{(ii)} are possible: $\eta(P_g)$ is decreasing when $\mathbb{F}_2(P_g)<0$ $\forall P_g\in(0,\alpha_1)$, $\eta(P_g)$ increasing when $\mathbb{F}_2(P_g)>0 \ \forall P_g\in (\alpha_1,\alpha_2)$ and $\eta(P_g)$ is decreasing when $\mathbb{F}_2(P_g)<0$ $\forall P_g\in(\alpha_2, \infty)$. Hence, $\eta(P_g)$ is not quasi-concave w.r.t $P_g$. \par
Hence, in case \textbf{(i)}, for a given $R_g$, it may be noted that EE is maximum when $\lim\limits_{P_g \to 0} \eta(P_g)$. Also, in case \textbf{(ii)}, the maximum EE is $max(\lim\limits_{P_g \to 0}\eta(P_g), \eta(\alpha_2))$. However, to satisfy the minimum rate and maximum outage probability constraints, specified in (\ref{m8}), the transmit power cannot be arbitrarily reduced, such that $P_g \to 0$. The CBS has to transmit with a minimum power $P_g^{min}$ to support the application with rate $R_g^{min}$ and outage $\mathscr{P}_g^{out^{(max)}}$.  
Hence, the optimal power $P_g^{\ast}$, $g\in 1,2, \cdots ,G$ is given as:
\begin{equation} \label{m15}
P_g^{\ast}=
\begin{cases}
      \alpha_2, \ \ \mathbb{F}_2(\beta)>0 \ \ \& \ \eta(P_g^{min})\leq \eta(\alpha_2)\\\
      P_g^{min}, \ \normalfont{otherwise}.
    \end{cases}
\end{equation}
Thus, $P_g^{\ast}$ can be obtained from (\ref{m15}) by evaluating $\alpha_2$ using bisection method. However, a specific case may occur when $P_g^{\ast}$ violates IPT constraint in (\ref{m7}), i.e $\alpha_2 > \frac {Q_m}{\lambda_{m,g}^{sp}}$. In such a situation, the optimal power, $P_g^{\ast}$ is obtained at $P_g^{max}=\frac {Q_m}{\lambda_{m,g}^{sp}}$.
\begin{figure*}[t]
\begin{subfigure}[t]{.33\textwidth}
  \includegraphics[width=\textwidth]{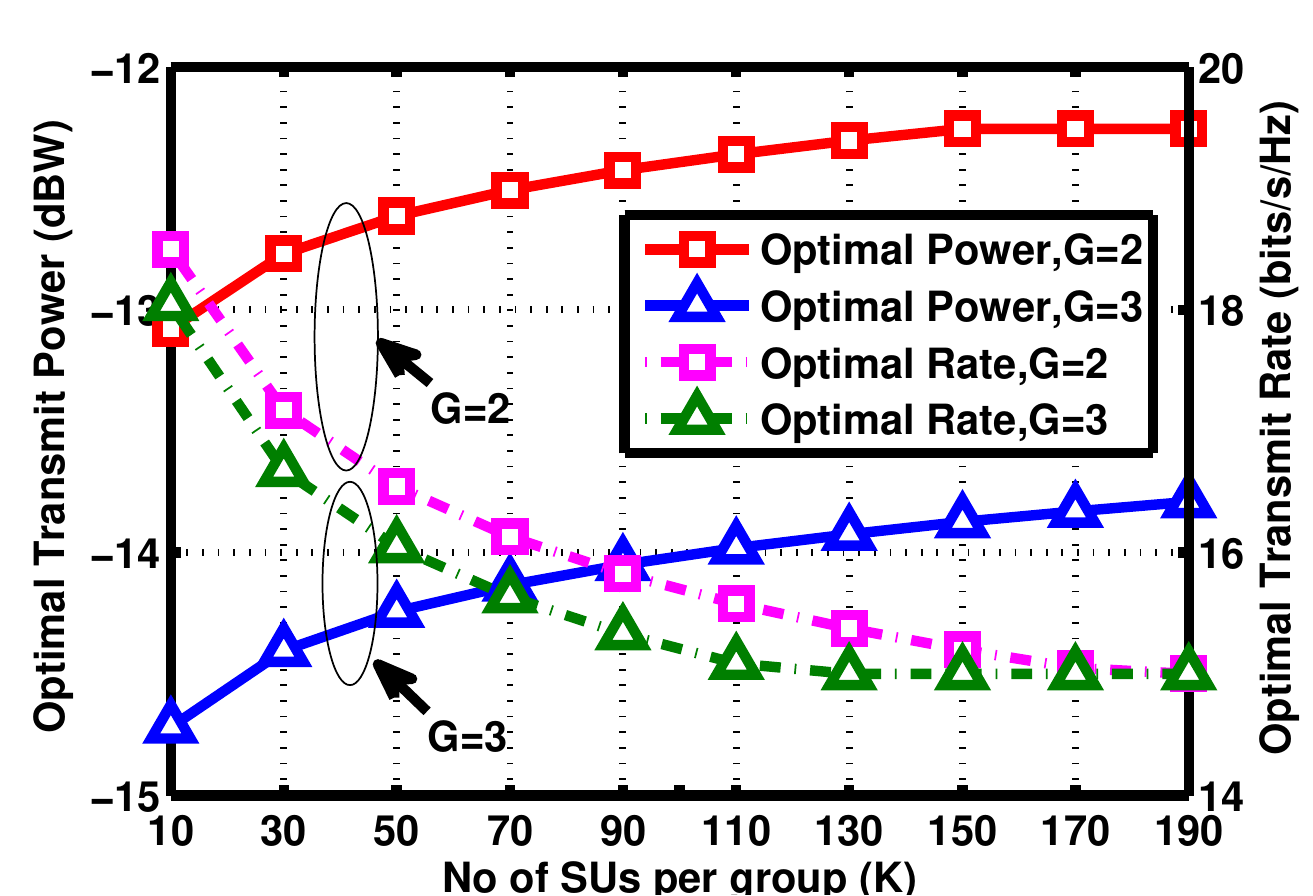}
  \caption{}
  \label{f2}
\end{subfigure}
\begin{subfigure}[t]{.33\textwidth}
  \includegraphics[width=\textwidth]{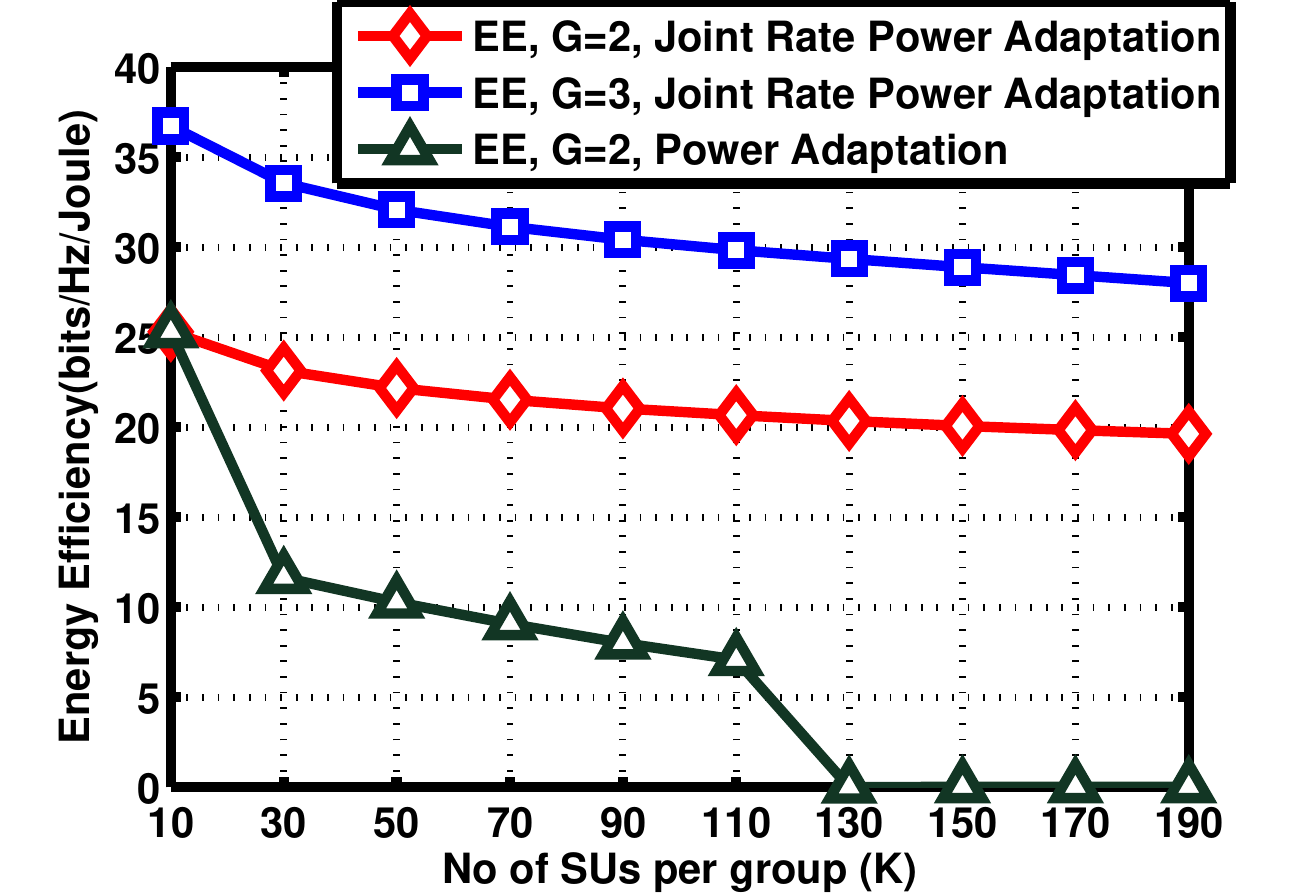}
 \caption{}
  \label{f3}
\end{subfigure}
\begin{subfigure}[t]{.33\textwidth}
  \includegraphics[width=\textwidth]{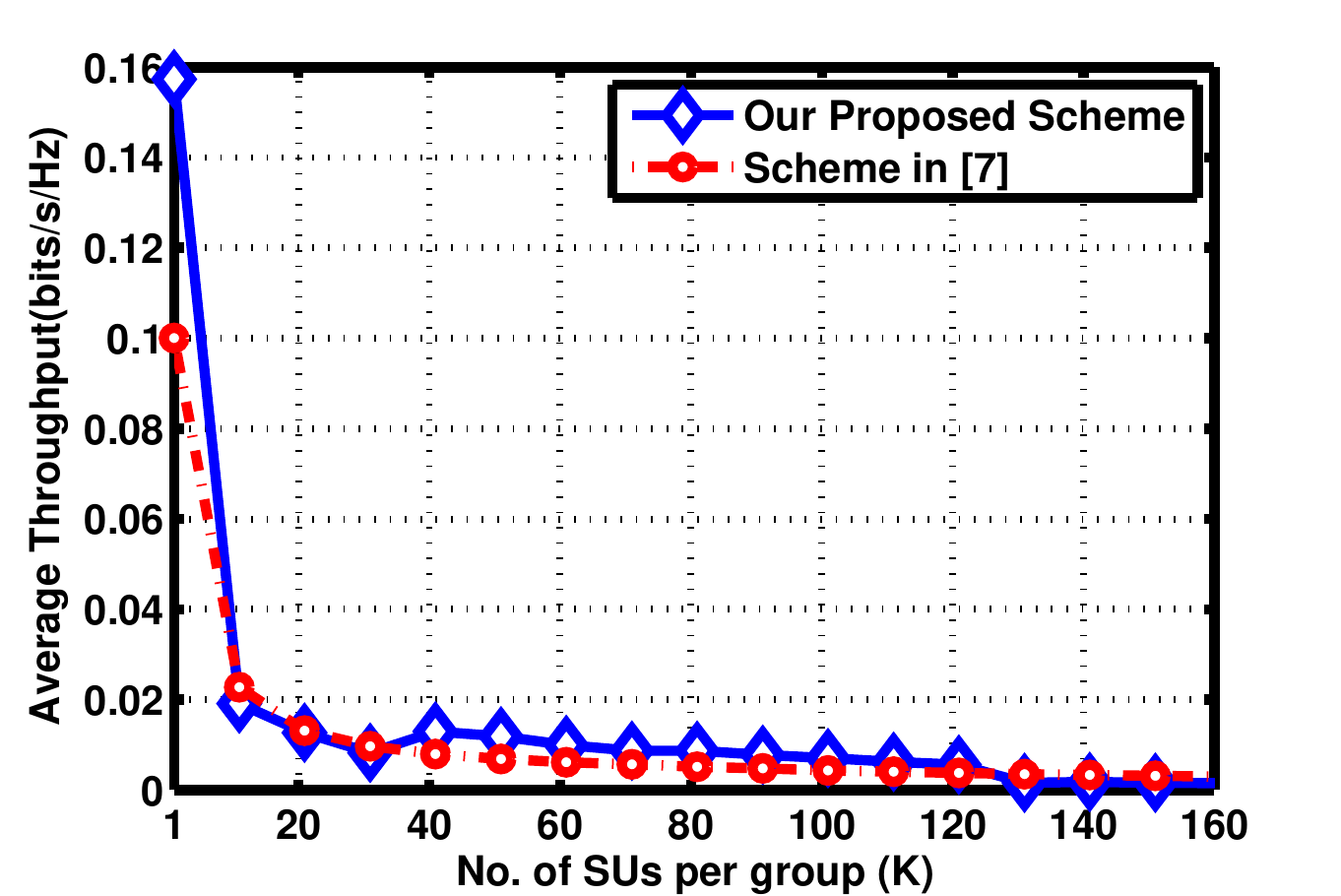}
  \caption{}
  \label{f4}
\end{subfigure}
\caption{Simulation Results: (a) Optimal Rate and Power vs. Number of SUs, (b) EE vs. Number of SUs, (c) Average group throughput vs. Number of SUs}
\end{figure*}
\subsection{Maximize $\eta$ w.r.t $R_g$} \label{s5}
Considering $\eta(P_g,R_g)$ as a function of $R_g$ only, and treating all other variables constants, the $1^{st}$ order derivative of (\ref{m6}) w.r.t $R_g$ is
\begin{equation}\label{m17}
\eta^{(1)}(R_g)=\frac{\mathbb{L}_1(R_g) \mathbb{L}_2(R_g)}{P_g+P_c}, \ \ \text{where}
\end{equation}
\begin{equation}\label{m18}
\mathbb{L}_1(R_g)=e^{-\varphi(2^{R_g}-1)} \ \ \text{where}
\end{equation}
\begin{equation} \label{m18a}
\mathbb{L}_2(R_g)=1-\varphi R_g2^{R_g}\ln 2
\end{equation}
where $\varphi =\frac{N_0|K_g|}{\lambda_{g}^{ss}P_g}$. It can be observed from above that $\mathbb{L}_2(R_g)$ determines the sign of $\eta^{(1)}(R_g)$ in (\ref{m17}). Hence, differentiating $\mathbb{L}_2(R_g)$  w.r.t $R_g$,
\begin{equation} \label{m19}
\mathbb{L}_2^{(1)}(R_g) = -\varphi 2^{R_g}\ln 2\left[ 1 + R_g 2^{R_g}\ln 2\right]
\end{equation}
Thus, from (\ref{m19}), it is clear that $\mathbb{L}_2(R_g)$ is strictly decreasing function in $R_g\in (0,\infty)$. Moreover, $\mathbb{L}_2(R_g)$ is continuous in $(0,\infty)$, while $\lim\limits_{R_g\to 0} \mathbb{L}_2(R_g)=1$ and $\lim\limits_{R_g\to \infty} \mathbb{L}_2(R_g)=-\infty$. Hence, a point $\delta$ exists such that $\mathbb{L}_2(\delta)=0$. From above, it is obvious that $\eta(R_g)$ is strictly increasing in $R_g\in (0, \delta)$ and strictly decreasing in $R_g \in (\delta, \infty)$. Therefore, $\eta^{(1)}(R_g)$ is a quasi-concave function of $R_g$.
The optimal target rate $R_g$ of CBS, for a given power $P_g$, $g\in 1,2, \cdots, G$ is given below:
\begin{equation} \label{m20}
R_g^{\ast}=\delta
\end{equation}
We use bisection method to evaluate $\delta$. If $R_g^{\ast}$ is beyond the specified rate limits in (\ref{m8}), then optimal rate is equal to either of the boundary values.
\subsection{Joint rate and power control Algorithm}
Energy Efficient joint rate and power control for \textbf{P1} is shown in \textbf{Algorithm 1}. For each multicast group, the optimal values of $\eta(P_g,R_g)$ w.r.t $P_g$ and $R_g$ are updated in a cyclic order at each iteration using block coordinate descent method \cite{bertsekas1999nonlinear}.
\subsubsection{Convergence}
Block coordinate method \cite{bertsekas1999nonlinear} can be applied to an optimization problem if the cost function and the constraints have a partially decomposable structure w.r.t each optimization variable as in (\ref{m8}). It is proved in \cite [proposition 2.7.1]{bertsekas1999nonlinear} that if the objective function is continuously differentiable in its domain and attains an unique optima for each design variable, this method converges to a stationary point. Also, in \cite[Section 6]{grippof1999globally}, it is proved that- "for two block decomposition, this method is globally convergent towards stationary point, even in the absence of convexity or uniqueness assumptions". Since $\eta(P_g,R_g)$ is continuously differentiable and has unique maxima w.r.t $P_g$ \& $R_g$ as proved in \ref{s4} \& \ref{s5}, the global maximum for $\eta$ can be attained by alternately updating optimal power ($P_g^{\ast}$) and optimal rate ($R_g^{\ast}$) for each group in every iteration.
\subsubsection{Complexity Analysis}
Nested bisection search is applied in (\ref{m13}) and (\ref{m14}) to find $P_g^{\ast}$, while $R_g^{\ast}$ is obtained using bisection search in (\ref{m18a}). The complexity of bisection method is $\mathcal{O}(\log_2K)$, where $K$ is the number of subintervals in the search range \cite{cormen2009}. The complexity of bisection method is $\mathcal{O}(\log_2K)$, where $K$ is the number of subintervals in the search range 
Hence, the complexity of \textbf{Algorithm 1} is $\mathcal{O}(G\log_2(K_P^2K_R))$, where $K_P$, $K_R$ are the number of subintervals in search ranges of $P_g$ and $R_g$ respectively.
\begin{algorithm} [h]
\begin{small}
\caption{Energy Efficient Joint Rate and Power Allocation}
\label{Al1}
\begin{algorithmic}[1]
\State{Initialize $P_g$  to a low value and $R_g$ to a high value ($R_g\leq R_g^{max}), \ \forall g\in 1,2, \cdots, G$};
\Repeat
\State{Update i=i+1};
\State{Input: $\left\lbrace P_g^{(i-1)}\right\rbrace _{g=1}^{G}, \left\lbrace R_g^{(i-1)}\right\rbrace _{g=1}^{G}$}
\For{$g=1$ to $G$}
\State {Solve (\ref{m15}) and update $P_g^{(i)}\leftarrow P_g^{\ast}$ taking $R_g^{(i)}=R_g^{(i-1)}$}
\State {Solve (\ref{m20}) and update $R_g^{(i)}\leftarrow R_g^{\ast}$} 
\EndFor
\Until{Convergence = true}
\If {$P_g^{(i)}$ satisfies equation (\ref{m7})} {$P_g^{\ast}\leftarrow P_g^{(i)}$}
\Else {} {$P_g^{*}\leftarrow \frac{Q_m}{\lambda_{m,g}^{sp}}$}
\EndIf
\If {$R_g^{(i)}> R_g^{max}$} {$R_g^{\ast}\leftarrow R_g^{max}$}
\ElsIf {$R_g^{(i)} \leq R_g^{max} \ \&\& \ R_g^{(i)} \geq R_g^{min}$} {$R_g^{\ast}\leftarrow R_g^{(i)}$}
\Else {} {$R_g^{\ast}\leftarrow R_g^{min}$}
\EndIf
\State{Output: $\left\lbrace P_g^{(\ast)}\right\rbrace _{g=1}^{G}, \left\lbrace R_g^{(\ast)}\right\rbrace _{g=1}^{G}$}
\end{algorithmic}
\end{small}
\end{algorithm}
\section{Simulation Results}
This section presents the results of performance evaluation of the proposed energy efficient joint power and rate allocation strategy.  Simulations are done in MATLAB\textsuperscript{\textregistered} taking the values of the various parameters as follows: the mean of channel power gain from  CBS to SUs, $\lambda_g^{ss}$ and CBS to PUs, $\lambda_{m,g}^{sp}$ are assumed to be 1, noise power, $N_0 = -90$ dB, circuit power, $P_c = 1$ dB and $\mathscr{P}_g^{out^{(max)}}=30\% $. For simplicity, 
$R_g^{max}$ and $R_g^{min}$  are taken equal for all groups, with values 18.5 bps/Hz and 15 bps/Hz respectively. The IPT constraint is considered identical for all PUs and set at $-12.3$ dB. To investigate the individual impact of multicast group size, $K_g$ and no. of simultaneously active groups $G$, we assume identical no. of SUs are present in each group.\par 
Fig. \ref{f2} shows the variations of optimal transmit power, $P_g^{\ast}$ and target rate, $R_g^{\ast}$ with the no. of SUs in each group, $K$ for two different no. of active multicast groups: $G=2$ and $G=3$.  As shown in Fig. \ref{f2}, for a both the values of $G$, as $K$ increases, the CBS increases its transmit power while reduces the rate for each group. For a fixed rate and power, larger $K$ causes increase in outage, $\mathscr{P}_{g}^{out}$ following (\ref{m3}), which in turn tends to lower EE as shown in (\ref{m5}). Thus, to maximize EE with increase in $K$, CBS raises $P_g$ until it reaches IPT constraint (\ref{m7}) and simultaneously lowers $R_g$, as long as the minimum rate constraint is satisfied. Interestingly, it may also be observed that for the same set of values of $K$ for each group, the presence of more no. of $G$ in the multicast system lowers the optimal transmit power and rate for each group.

Fig. \ref{f3} shows the variation of optimal EE with $K$ and $G$. It is seen that for a specific $G$, EE gradually decreases as $K$ is augmented. As depicted in Fig. \ref{f2}, when $K$ increases, CBS increases $P_g^{\ast}$ and lowers $R_g^{\ast}$ and this reduces EE. It may also be noted EE is higher for multicast system having higher $G$. This results from the fact that optimal $P_g$ is $12\%$ lower while $R_g$ reduces by only $3\%$, when $G$ is increased from $2$ to $3$. \par 
In Fig. \ref{f3}, we also compare our proposed scheme with EE maximization using transmit power adaptation only, for rate $R_g=R_g^{max}$ and $G=2$. We observe that our scheme outperforms the other scheme. 
It may be noted that as the number of SUs in each multicast group increases, the outage probability becomes higher, following (\ref{m3}). Hence, CBS requires more transmit power to maximize EE, while supporting the given target rate. However, power cannot be increased further once its meets the corresponding IPT constraint. Thus EE, for only power adaptation scheme, is significantly lower than that in our proposed scheme. This also advocates the efficacy of cross layer approach. \par
In Fig. \ref{f4}, the average throughput (\ref{m5}) of each multicast group in our proposed scheme is compared with another joint rate and power control scheme for throughput maximization in CR as proposed in \cite{xu2013joint}. The simluation parameters are same as in \cite{xu2013joint} and $P_c=0$. For proper comparison, the SU's service outage constraint, $\epsilon$ in \cite{xu2013joint} is taken as $0$. It is observed that the performance of both these schemes are quite similar.
\vskip -2mm
\section{Conclusion}
This paper addresses the problem of energy efficient multicasting in CR considering underlay mode of operation. Our proposed solution uses joint rate and power adaptation to achieve better EE than that of transmit power adaptation only. Further, simulation results show that system EE improves with the increase in number of simultaneously active multicast groups, provided necessary number of PU channels are available. This may also considered beneficial from the commercial perspective of network operation.
\appendices
\section{Proof of Lemma \ref{L1}}
Let $h_{g,1}, h_{g,2},\ldots, h_{g,|K_g|}$ be the channel power gains of SUs of group $g \in 1,2, \cdots, G$, such that $ h_{g,k}^{ss} \sim \frac{1}{\lambda_g^{ss}}e^{-\frac{h_{g,k}^{ss}}{\lambda_g^{ss}}} \ , \ k\in K_g $.
\begin{align*}
r_{g,k}&=\log_2\left(1+\frac{h_{g,k}^{ss} P_{g}}{N_0}\right)
\end{align*}
Since $h_{g,k}$ follows exponential distribution, CDF of $r_{g,k}$ is expressed as
\begin{align*}
Pr\left( r_{g,k} \leq R_g \right)  =  1-e^{-\frac{No}{\lambda_{g}^{ss}P_g}(2^{R_g}-1)}
\end{align*}
Since all SUs in a group are i.i.d, the outage probability of multicast group $g$  is given by
\begin{align*}
\mathscr{P}_{g}^{out} &= 1- Pr\left\lbrace \min \left( r_{g,1}, r_{g,2}, \ldots, r_{g,{|K_g|}} \right) > R_g \right\rbrace \\
&=1-\left\lbrace Pr\left( r_{g,1} > R_g\right)Pr\left(r_{g,2} > R_g \right)\ldots Pr\left(r_{g,k} > R_g \right)\right\rbrace \\
&=1-\left[ Pr\left(r_{g,k} > R_g \right)\right] ^{|K_g|} \\
&=1-e^{-\frac{No|K_g|}{\lambda_{g}^{ss}P_g}(2^{R_g}-1)}
\end{align*}

\bibliographystyle{IEEEtran}
\bibliography{IEEEabrv,ref}
\end{document}